\documentclass[journal]{IEEEtranTCOM}
\normalsize
\usepackage[pdftex]{graphicx}
  \usepackage{epstopdf}
  \DeclareGraphicsExtensions{.pdf,.jpeg,.png,.eps}
\usepackage{color,balance}
\usepackage[cmex10]{amsmath}
\usepackage{amssymb}
\usepackage{algorithmic}
\usepackage{array}
\usepackage{mdwmath}
\usepackage{amsthm}
\usepackage{multirow}
\usepackage{booktabs}
\usepackage{comment}
\usepackage{graphicx}
\usepackage{textgreek}
\usepackage{float}
\usepackage{cite}
\usepackage{amsmath}
\usepackage{bm}
\usepackage{xr}
\usepackage{diagbox}

\usepackage{mdwtab}
\usepackage{tabularx}
\usepackage{xcolor}
\usepackage{xpatch}
\usepackage{adjustbox}
\usepackage{balance}
\usepackage{cite}
\makeatother

\allowdisplaybreaks
\newtheoremstyle{mystyle}
  {3pt}
  {3pt}
  {\itshape} 
  {\parindent}
  {\bfseries}
  {\upshape{:}}
  {.5em}
  {}

\usepackage{color,soul}
\usepackage{setspace}
\theoremstyle{mystyle}

\theoremstyle{mystyle}  

\theoremstyle{mystyle}

\usepackage[british,UKenglish]{babel}
\usepackage[font={small}]{caption}
\graphicspath{{./figures/}}

\usepackage{tabularx}
\usepackage{booktabs}
\usepackage{array}

\begin{document}

\title{Enhancing Reliability in Federated mmWave Networks: A Practical and Scalable Solution using Radar-Aided Dynamic Blockage Recognition}

\author{Mohammad Al-Quraan,~\IEEEmembership{Graduate Student Member,~IEEE,} Ahmed Zoha,~\IEEEmembership{Senior Member,~IEEE,}  Anthony Centeno, Haythem Bany Salameh,~\IEEEmembership{Senior Member,~IEEE,} Sami Muhaidat,~\IEEEmembership{Senior Member,~IEEE,} Muhammad Ali Imran,~\IEEEmembership{Fellow,~IEEE} and  Lina Mohjazi,~\IEEEmembership{Senior Member,~IEEE.}

\thanks{M. Al-Quraan, A. Zoha, A. Centeno, M. A. Imran, and L. Mohjazi are with the James Watt School of Engineering, University of Glasgow, Glasgow, G12 8QQ, UK. (e-mail: \{m.alquraan.1\}@research.gla.ac.uk, \{Ahmed.Zoha, Anthony.Centeno, Muhammad.Imran, Lina.Mohjazi\}@glasgow.ac.uk).}

\thanks{H. Bany Salameh is with the Networks, Communication Engineering Department, Al Ain University, Al Ain 64141, UAE, and also with the Telecommunications Engineering Department, Yarmouk University, Irbid
21163, Jordan (e-mail: haythem@email.arizona.edu).}

\thanks{S. Muhaidat is with the KU Center for Cyber-Physical Systems, Department of Electrical Engineering and Computer Science, Khalifa University, Abu Dhabi 127788, UAE, and with the Department of Systems and Computer Engineering, Carleton University, Ottawa, ON K1S 5B6, Canada (e-mail: muhaidat@ieee.org).}

}

\maketitle
\markboth{}{}
\begin{abstract}
This article introduces a new method to improve the dependability of millimeter-wave (mmWave) and terahertz (THz) network services in dynamic outdoor environments. In these settings, line-of-sight (LoS) connections are easily interrupted by moving obstacles like humans and vehicles. The proposed approach, coined as Radar-aided Dynamic blockage Recognition (RaDaR), leverages radar measurements and federated learning (FL) to train a dual-output neural network (NN) model capable of simultaneously predicting blockage status and time. This enables determining the optimal point for proactive handover (PHO) or beam switching, thereby reducing the latency introduced by 5G new radio procedures and ensuring high quality of experience (QoE). The framework employs radar sensors to monitor and track objects movement, generating range-angle and range-velocity maps that are useful for scene analysis and predictions. Moreover, FL provides additional benefits such as privacy protection, scalability, and knowledge sharing. The framework is assessed using an extensive real-world dataset comprising mmWave channel information and radar data. The evaluation results show that RaDaR substantially enhances network reliability, achieving an average success rate of 94\% for PHO compared to existing reactive HO procedures that lack proactive blockage prediction. Additionally, RaDaR maintains a superior QoE by ensuring sustained high throughput levels and minimising PHO latency.

\end{abstract}


\section{Introduction}
Millimetre-wave (mmWave) and terahertz (THz) technologies are anticipated to revolutionise wireless communication systems due to their large available bandwidths that can offer multi-Gbit/s data rates \cite{[1.30]}. However, high-frequency bands suffer significant challenges stemming from their electromagnetic properties, including propagation loss, atmospheric attenuation, and susceptibility to blockages. Therefore, massive multiple-input multiple-output (MIMO) and beamforming techniques are becoming indispensable enablers in next-generation wireless networks. MIMO can compensate for the propagation and attenuation losses using large antenna arrays to generate narrow line-of-sight (LoS) beams, which improve the quality of the received signal. Nevertheless, LoS links experience rapid and temporary fluctuations in the received signal strength (RSS) when obstructed by obstacles, especially in dynamic environments like urban areas. This may lead to frequent handovers (HOs), which negatively impact the network latency and reliability. Thus, it is crucial to comprehensively address these challenges to fully exploit the potentials of mmWave and THz bands and develop reliable wireless communication systems.

Recently, several mechanisms and methods have been proposed in the literature to alleviate the impact of beam blockage. The traditional approach entails implementing a multi-connectivity mechanism \cite{[1.31]}, wherein user equipment (UE) is connected to multiple base stations (BSs) simultaneously. As a result, in the event of a blockage, the UE can still be served by other intact links, increasing the likelihood of maintaining service continuity. Other techniques leverage dual-band operation, i.e., sub-6GHz and super-6GHz (24-86GHz), to identify blockages \cite{[1.13]}. These methods capitalise on the fluctuations in lower-frequency signals caused by signal diffraction and employ them as an early warning of potential link blockages at higher frequencies. Instead of complicating the communication system, recent studies suggested monitoring the behaviour of non-LoS components of the mmWave links to predict whether the primary LoS component will experience a blockage \cite{[1.32]}. However, these methods are typically reactive in their response to blockage events.  This is due to their inadequate ability to predict beam blockages with a sufficient prediction interval that allows the network to implement preemptive measures. Hence, service interruption occurs with each blockage event.

One promising solution for addressing LoS link disconnection issues is to proactively predict blockages before they happen, enabling the network to make timely decisions, such as beam switching or handover. To accomplish this, one possible solution is to leverage the advancements in machine learning (ML) and utilize relevant information about the environment surrounding the LoS links to anticipate future obstacles, thereby acquiring proactive capabilities. Sensing modalities, including vision, light detection and ranging (LiDAR), and radar sensors, are used to increase the network's awareness of the surrounding environment. Recent studies explored the use of network-side information gathered from these sensors to address challenging problems associated with mmWave networks, such as user positioning, beam selection, and blockage prediction. Early works in \cite{[1.10]} have shown promising results when leveraging camera imagery information and deep learning (DL) models \cite{[1.40]} to anticipate blockage events before they happen. However, deploying vision sensors may not always be feasible due to regulatory and privacy concerns. Moreover, image quality may be affected by low-light and adverse weather conditions. Another line of work considers LiDAR sensors to aid the operation of mmWave networks by utilising point clouds captured from the served environment \cite{[1.5]}. Similarly, such approaches are affected by weather conditions, and their practicality is limited to specific scenarios.

The aforementioned limitations have spurred a new research direction that involves the adoption of radar technology for predicting link blockages. This is achieved by detecting obstacles through radar fingerprints. The reasons for this shift to radar-based solutions are due to their low cost, ability to capture useful object features like range, velocity, and direction. In addition, radar technology offers lower privacy risk and, most importantly, enables low-latency transmissions as it operates at high-frequency bands. To date, few studies have considered the use of radars to address beam blockages problem in high-frequency networks \cite{[1.18],[1.17]}. However, these studies are preliminary and restricted to specific scenarios. Furthermore, they did not adequately consider the dimensions of obstacles, which are crucial factors in determining the potential blockage of an LoS link. Additionally, existing studies overlooked the significance of the height of both the BS and the UE, which significantly influence the status of the mmWave channel. Consequently, further research is required to fully explore the potential of using radars to proactively predict blockages and achieve promising reliability and latency gains for the practical realisation of next-generation networks, which is the focus of this paper.

Specifically, in this work, we introduce a novel framework named radar-aided dynamic blockages recognition (RaDaR) that handles the challenge of frequent beam blockages in high-frequency outdoor networks. RaDaR employs radar sensors to enhance the network’s situational awareness by monitoring and tracking the movement of objects to generate range-angle and range-velocity maps. These maps are useful for analysing the scene and making accurate blockage predictions.  Unlike previous related works, RaDaR leverages radar information to predict the height of objects, a critical factor in determining whether an object will obstruct the LoS link or not. Additionally, the framework's end-to-end execution time is measured to provide the network with a proactiveness merit in predicting blockages and performing HO, to avoid link interruption and ensure high QoE. In addition, RaDaR incorporates the FL training mechanism, which brings three important benefits: scalability, knowledge sharing, and resource efficiency.  As a result, RaDaR supports network expansion to new development areas and consolidates knowledge from multiple SBSs by performing distributed learning rather than centralised learning. This consequently reduces the overhead on the network's transmission resources. Benefiting from the real-world DeepSense dataset \cite{[1.19]}, a set of non-independent and identically distributed (non-IID) features is extracted to collaboratively train a dual-output NN model offline using a group of SBSs. The trained model is then used online at each SBS to predict the occurrence of a blockage event and the remaining time before the obstacle obstructs the link. With this information, RaDaR can preemptively decide the optimal instant to switch the beam or initiate proactive HO (PHO) to maintain the users’ QoE as high as possible. Our contributions can be summarised as follows.
\begin{itemize}
    \item We present a novel framework, called RaDaR, which aims to improve the reliability of federated mmWave networks by integrating radars for the anticipation of LoS link blockages while considering latency and QoE metrics. 

    \item We utilise the FL algorithm to perform collaborative model training at each SBS by using information acquired from radar placed at the top of the SBS. FL provides the framework with vital features, including scalability, knowledge sharing, and conserving network resources.  

    \item We employ the large-scale real-world DeepSense dataset to evaluate the effectiveness of our proposed framework. Specifically, we augment scenario 30 and create diverse environments that reflect practical scenarios.
\end{itemize}

The rest of this paper is structured as follows. Section \ref{RW} discusses the related work. Section \ref{system_model} presents the system, channel, and blockage models. Section \ref{PF} provides a formal description of the beam blockage and PHO problem that is addressed in this paper. The proposed framework is discussed in detail in Section \ref{Framework}. Section \ref{PEval} introduces the experimental and simulation setup, along with the main results of that evaluation. Finally, Section \ref{Conclusion} gives concluding remarks.

\section{Related Work}\label{RW}
In this section, we review the state-of-the-art studies that have proposed solutions to mitigate the issue of beam blockage in high-frequency communication systems. These works can be categorised into two main strategies in terms of data acquisition for training the ML models: (i) wireless information-based approaches \cite{[1.12],[1.9],[1.13],[1.3]} and (ii) sensing information-based approaches \cite{[1.5],[1.1],[1.10],[1.15],[1.16]}. Table \ref{tab:Comparison} briefly outlines the key techniques, including their advantages and limitations.


\begin{table*}
\fontsize{6.5pt}{6.5pt}\selectfont
\caption{Existing research about the challenge of beam blockage in mmWave networks.}
\centering
\begin{tabular}{llllll}
\toprule
\multicolumn{1}{c}{\textbf{Ref.}} &
  \multicolumn{1}{c}{\textbf{Date}} &
  \multicolumn{1}{c}{\textbf{Algorithm}} &
  \multicolumn{1}{c}{\textbf{\begin{tabular}[c]{@{}c@{}}Indoor or \\ Outdoor?\end{tabular}}} &
  \multicolumn{1}{c}{\textbf{Advantages}} &
  \multicolumn{1}{c}{\textbf{Limitations}} \\ \hline\hline
  \\[-0.6em]
\cite{[1.18]} &
  Oct. 2016 &
  Radar-based mmWave medium access (RadMAC). &
  Indoor &
  \begin{tabular}[c]{@{}l@{}}RadMAC can significantly enhance throughput\\ and link stability.\end{tabular} &
  \begin{tabular}[c]{@{}l@{}}No consideration for the effect of the BS/UE/object\\ heights.\end{tabular} \\ \hline
  \\[-0.6em]
\cite{[1.12]} &
  June 2020 &
  \begin{tabular}[c]{@{}l@{}}Multi-directional links quality prediction using an\\ LSTM model.\end{tabular} &
  Indoor &
  \begin{tabular}[c]{@{}l@{}}Predcting the quality of several links within a \\single or multiple cells.\end{tabular} &
  \begin{tabular}[c]{@{}l@{}}Insufficient time for optimising resources to mitigate\\ potential link failures.\end{tabular} \\ \hline
  \\[-0.6em]
\cite{[1.13]} &
  Oct. 2020 &
  \begin{tabular}[c]{@{}l@{}}DL-based blockage avoidance in IRS-assisted \\ mmWave networks.\end{tabular} &
  Outdoor &
  \begin{tabular}[c]{@{}l@{}}High efficient beam management with reduced\\ frequent handover.\end{tabular} &
  Increased network design and operation complexity. \\ \hline
  \\[-0.6em]
\cite{[1.10]} &
  June 2021 &
  \begin{tabular}[c]{@{}l@{}}CNN and RNN models trained using visual and\\ wireless information.\end{tabular} &
  Outdoor &
  \begin{tabular}[c]{@{}l@{}}CV is employed to foster an awareness of the\\ surrounding environment.\end{tabular} &
  \begin{tabular}[c]{@{}l@{}}Privacy concerns, challenges in low-light and bad\\ weather conditions, no target user identification.\end{tabular} \\ \hline
  \\[-0.6em]
\cite{[1.9]} &
  Dec. 2021 &
  Meta-learning based RNN for blockage prediction. &
  Indoor &
  Less data is required to train ML models. &
  \begin{tabular}[c]{@{}l@{}}Risk of overfitting, challenge for generalisation to\\ unseen data.\end{tabular} \\ \hline
  \\[-0.6em]
\cite{[1.17]} &
  May 2022 &
  \begin{tabular}[c]{@{}l@{}}CNN and LSTM blockage prediction based on\\ radar data.\end{tabular} &
  Outdoor &
  \begin{tabular}[c]{@{}l@{}}Predicting link disruption up to one second\\ ahead.\end{tabular} &
  \begin{tabular}[c]{@{}l@{}}The study is limited to a very specific scenario, not\\ considering the effect of the BS/UE/vehicle heights.\end{tabular} \\ \hline
  \\[-0.6em]
\cite{[1.3]} &
  June 2022 &
  \begin{tabular}[c]{@{}l@{}}LSTM-based blockage prediction using signal\\ diffraction characterestics.\end{tabular} &
  Both &
  \begin{tabular}[c]{@{}l@{}}Generalisation from outdoor to indoor\\ scenarios.\end{tabular} &
  An increase in both system complexity and cost. \\ \hline
  \\[-0.6em]
\cite{[1.6]} &
  June 2022 &
  \begin{tabular}[c]{@{}l@{}}DNN model trained on beam measurement\\ reports.\end{tabular} &
  Indoor &
  \begin{tabular}[c]{@{}l@{}}Rely exclusively on the mmWave channel\\ measurements.\end{tabular} &
  Evaluations based on limited obstacles speed \\ \hline
  \\[-0.6em]
\cite{[1.15]} &
  Oct. 2022 &
  \begin{tabular}[c]{@{}l@{}}Latency-aware vision-aided proactive blockage\\ prediction.\end{tabular} &
  Outdoor &
  Reduced network latency and high QoE. &
  \begin{tabular}[c]{@{}l@{}}Privacy concerns, limitations in low-light and bad\\ weather conditions.\end{tabular} \\ \hline
  \\[-0.6em]
\cite{[1.1]} &
  Jan. 2023 &
  Channel-oriented semantic communications. &
  Outdoor &
  Enhanced mmWave system reliability. &
  \begin{tabular}[c]{@{}l@{}}Challenges in low-light and bad weather conditions,\\ no target user identification.\end{tabular} \\ \hline
  \\[-0.6em]
\cite{[1.5]} &
  Jan. 2023 &
  \begin{tabular}[c]{@{}l@{}}3DCNN and GBRT models trained using\\ LiDAR information.\end{tabular} &
  Indoor &
  \begin{tabular}[c]{@{}l@{}}Anticipate attenuation in link quality up to \\ 1000 ms ahead.\end{tabular} &
  \begin{tabular}[c]{@{}l@{}}Limited coverage area, less effective in bad weather\\ conditions.\end{tabular} \\ \hline
  \\[-0.6em]
\cite{[1.16]} &
  Mar. 2023 &
  \begin{tabular}[c]{@{}l@{}}FL-based latency-aware dynamic blockage\\ prediction.\end{tabular} &
  Outdoor &
  \begin{tabular}[c]{@{}l@{}}FL algorithm is employed, and latency aware\\ proactive HO triggering.\end{tabular} &
  Challenges in low-light and bad weather conditions. \\ \bottomrule
\end{tabular}
\label{tab:Comparison}
\end{table*}

\subsubsection{Wireless information-based approaches}
Those studies are further classified into three distinct groups based on the targeted environment: indoor, outdoor, or both. For example, the work in \cite{[1.12]} proposed utilising a long short-term memory (LSTM) network to predict the fluctuations in received power for one or more links in the next time instant. Others relied on the recurrent neural network (RNN) and meta-learning concept to predict blockages in indoor industrial environments \cite{[1.9]}. Considering outdoor methodologies, the study in \cite{[1.13]} focused on controlling the mmWave propagation channel by adopting intelligent reflective surfaces (IRSs) to avoid blockages. In contrast, the authors in \cite{[1.3]} employed the diffraction characteristics of the signals in sub-6GHz uplinks to predict mmWave blockages by training an LSTM model.

\subsubsection{Sensing information-based approaches}
Similarly, while some of the sensing-aided works are tailored for indoor networks \cite{[1.5]}, the majority of them are designed for outdoor systems \cite{[1.1],[1.10],[1.15],[1.16]}. The work in \cite{[1.5]} relied on LiDAR point cloud and ML to perform proactive predictions about mmWave channel received power. On the other hand, the concept of semantic communication was utilised in \cite{[1.1]} to anticipate blockages and aid the operation of the network. Recently, a new paradigm called vision-aided wireless communication has emerged with the aim of leveraging vision sensors to develop an awareness of the surrounding environment and facilitate the operations of mmWave networks. This approach has been investigated in several previous works, including \cite{[1.10],[1.15],[1.16]}. In \cite{[1.10]}, a fusion of RGB images and beamforming vectors was employed to train DL models and predict LoS link blockages before they occur. In addition, the authors in \cite{[1.15],[1.16]} proposed frameworks that employs computer vision (CV) and multi-output NN models to proactively predict the blockages and estimate the time until the user gets blocked.

Radar-based blockage prediction is a nascent research direction inspired by the advantageous features of radar technology. To date, only a limited number of studies have explored the effectiveness of this approach \cite{[1.18],[1.17]}. For instance, the work in \cite{[1.18]} is utilising radar sensors to enhance the reliability of mmWave systems in indoor environments. The study proposed a concept called radar-based medium access (RadMAC), which leverages reflected radar signals to enable intelligent beam-steering decisions based on blockage prediction and avoidance. Likewise, in \cite{[1.17]}, it was demonstrated that mmWave BSs can be integrated with frequency-modulated continuous-wave (FMCW) radars to obtain valuable information such as range and velocity that aid in predicting network obstacles. Nevertheless, these studies adopted a generalised blocking assumption that does not account for the dimensions of the detected objects, as well as the height of SBSs, UEs, or objects. As a result, such assumptions may provide inaccurate information on the disruption of the wireless link.

\section{System Model}\label{system_model}
\begin{figure}
\centering
\includegraphics[scale=0.42]{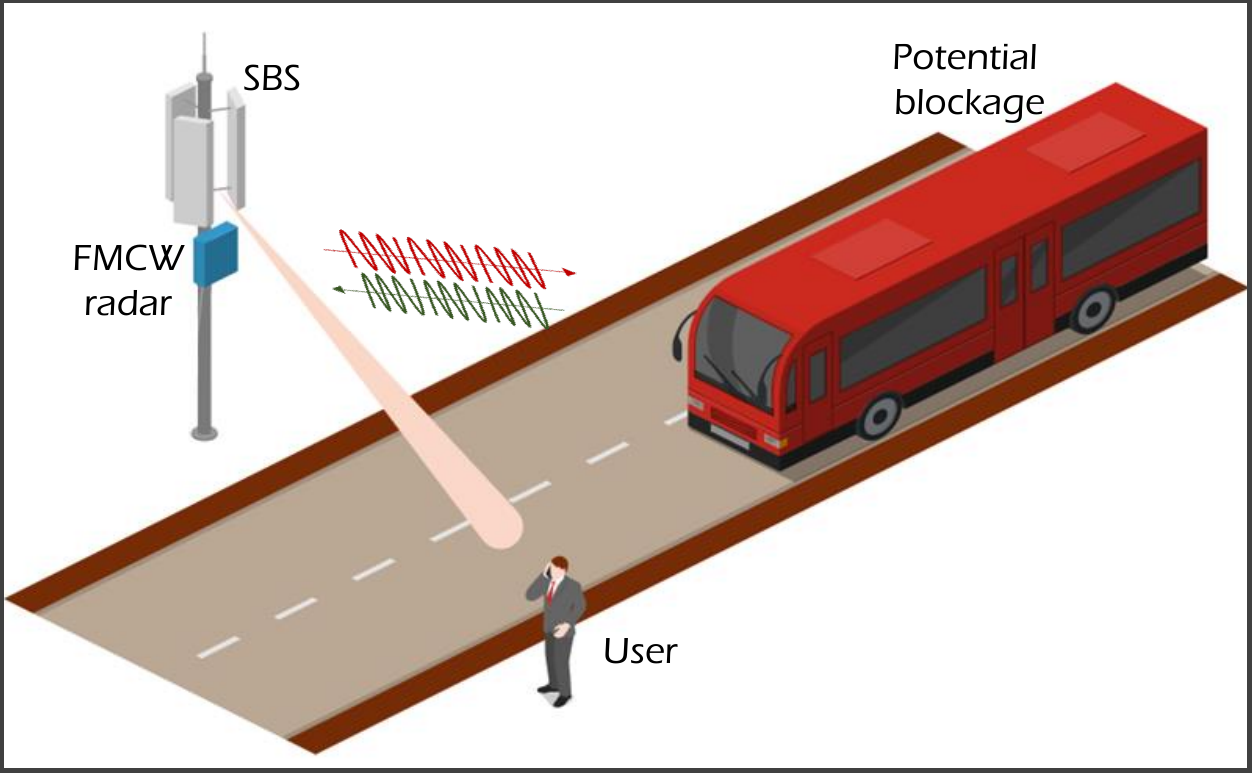}
\caption{The proposed system model of the RaDaR framework.}
\label{SM}
\end{figure}

The considered system comprises several SBSs and stationary\footnote{The term ‘stationary’ is used to indicate that the user will remain within the effective coverage range of the beam.} users in a vehicular environment, as shown in Fig. \ref{SM}. The SBSs and users are equipped with a global positioning system (GPS) that is supported by a real-time kinematic network. This enables accurate three-dimensional geolocation with sub-centimetre precision \cite{[1.33]}. Moreover, the SBS is equipped with two primary components: (i) a phased array antenna that produces LoS beams to serve the user and (ii) an FMCW radar mounted at the SBS to detect and track mobile objects in the operating vicinity. In the following subsections, we explain the signalling models for the network and radar components.

\subsection{Channel and Blockage Models}
The considered system operates at 60 GHz using orthogonal frequency division multiplexing (OFDM) with $K$ subcarriers and a cyclic prefix of length $Q$. Each SBS is equipped with a uniform linear array (ULA) consisting of $M$ antenna elements. These elements are utilised to produce narrow directive beams that maximise the receive beamforming gain at single-antenna UEs. Moreover, we assume that each SBS has a predefined beamforming codebook $\mathcal{F}=\left\{\mathbf{f}_{i}\right\}_{i=1}^{B}$, $\mathbf{f}_{i} \in \mathbb{C}^{M \times 1}$, where $B$ denotes the total number of beams in the codebook. Any beamforming vector can be represented as:
\begin{equation}
\mathbf{f}_{i}=\frac{1}{\sqrt{M}}\left[1\ e^{j \frac{2 \pi}{\lambda} d \sin \left(\psi_{i}\right)}\ \ldots\ e^{j \frac{2 \pi}{\lambda} (M-1) d \sin \left(\psi_{i}\right)}\right]^{T},
\end{equation}
where $\frac{1}{\sqrt{M}}$ represents the normalisation factor, $d$ denotes the distance between adjacent antenna elements, $\lambda$ is the wavelength of the carrier frequency, $\psi_{i} \in \{\frac{2 \pi {i}}{{B}}\}_{i=0}^{B-1}$ is the steering angle, and $T$ indicates the transpose notation.

The objective of the network is to determine the optimal beam $\mathbf{f}^{\star}$ that yields the highest RSS at the UE. To accomplish this, the SBS acquires a pilot message from the UE and uses it to train all the beams in the codebook to identify the best beam. Once selected, the received signal at the UE side on the $k$th subcarrier can be represented as:
\begin{equation}
y_{ k}=\sqrt{\mathcal{E}_k } \mathbf{h}_{k}^{H} \mathbf{f}^{\star} s_{k} + n_{k},
\end{equation}
where $\sqrt{\mathcal{E}}$ is the transmitter gain, $\mathbf{h} \in \mathbb{C}^{M \times K}$ indicates the narrow band channel between the SBS and the UE, $(\cdot)^{H}$ denotes the Hermitian transpose, $s$ is the transmitted data symbol, and $n \sim \mathcal{C N}\left(0, \sigma^2\right)$ is the additive white Gaussian noise with zero mean and $\sigma^2$ variance.

Assuming that the multi-path components of the signal arrive at the receiver through $P$ distinct paths. The shortest path is the LoS path, which is the path of interest in our study and is denoted by $p^{\star}$, and the other paths represent the NLoS components. Hence, the channel between the transmitter and the receiver can be mathematically expressed as:
\begin{equation}
\mathbf{h}_k =\mathbf{h}_k^{\mathrm{LoS}}+\mathbf{h}_k^{\mathrm{NLoS}},
\end{equation}
where the LoS  and NLoS channels are given respectively as follows \cite{[1.35]}:
\begin{equation}
\mathbf{h}_{k}^{\mathrm{LoS}}=\sum_{q=0}^{Q-1} \alpha_{p^{\star}} e^{-\mathrm{j} \frac{2 \pi k}{K} q} \ell\left(q T_{s}-\tau_{p^{\star}}\right) \mathbf{a}\left(\theta_{p^{\star}}, \phi_{p^{\star}}\right),
\end{equation}

\begin{equation}
\mathbf{h}_{k}^{\mathrm{NLoS}}=\sum_{q=0}^{Q-1} \sum_{p=1}^{P\backslash\{p^{\star}\}} \alpha_{p} e^{-\mathrm{j} \frac{2 \pi k}{K} q} \ell\left(q T_{s}-\tau_{p}\right) \mathbf{a}\left(\theta_{p}, \phi_{p}\right),
\end{equation}
where $\alpha_{p}$, $\tau_{p}$, $\theta_{p}$, $\phi_{p}$ are the gain, delay, azimuth, and elevation angles of the arrival of path $p$, respectively. $T_s$ is used to denote the sampling time, and $\mathbf{a}$ is the ULA steering vector \cite{[1.35]}.

This study employs a blockage fading channel model to include the impact of blockages on the communication system. Let $b$ denote the LoS blockage status and is defined as:
\begin{equation}
b[t]=\left\{\begin{array}{l}
1, \text{ \small LoS path is blocked} \\
0, \text{ \small LoS path is not blocked}.
\end{array}\right.
\end{equation}
Therefore, the mmWave channel for any subcarrier $k$ at the time instant $t \in \mathbb{Z}^{+}$ can be updated as follows:
\begin{equation}
\mathbf{h}_k[t] =(1-b[t]) \hspace{0.1cm}\mathbf{h}_k^{\mathrm{LOS}}[t]+\mathbf{h}_k^{\mathrm{NLOS}}[t].
\end{equation}

\noindent It is noteworthy that high-frequency communication systems typically have a limited number of NLoS links. The channel gains for those NLoS links are significantly smaller compared to those of the LoS links, even in the presence of blockages \cite{[1.20]}. Consequently, it is reasonable to disregard the NLoS components and concentrate on the LoS links.

\subsection{Radar Model}
In our system, an FMCW radar is installed in each SBS to obtain measurements of the surrounding environment and leverage them to develop a proactive mechanism for predicting potential network blockages. In each measurement, the radar transmits a frame of $L$ frequency-modulated chirps that represent continuous waves of radio signals separated by pause time $\tau_p$. Each chirp (a.k.a. ramp) has a linearly varying frequency that starts from $f_c$ and ends with $f_c+mt$, given as:
\begin{equation}
X_{chirp}(t)=
A_t \exp \left(j\left(\left(2 \pi f_c t+\pi m t^2\right)\right)\right)
, 0\leq t \leq	\tau_c
\end{equation}
where $A_t$ denotes the transmitter gain, $m=B/\tau_c$ is the slope of the chirp signal, which has bandwidth $B$ and duration $\tau_c$. Upon transmitting all chirps, the radar system remains inactive until the initiation of a new frame. However, during the frame time, the radar receives the signals that are reflected by the target objects. The received signals are subsequently directed to a quadrature mixer, which combines the transmitted and reflected chirps to generate in-phase and quadrature components. The mixed signals are then processed by a low pass filter to produce intermediate frequency (IF) signals. The IF signal captures the variations in frequency (a.k.a. beat frequency) and the phase between the transmitted and reflected signals. The IF signal can be mathematically represented as \cite{[1.41]}:
\begin{equation}
Y(t)=A_t A_r \exp \left(j\left(2 \pi m \tau_{rt} t + 2 \pi f_c \tau_{rt}\right)\right)
\label{IF}
\end{equation}
where $A_r$ is the receiver gain, $\tau_{rt} = 2r/c$ is the round-trip delay of the radar signal reflected from the object, which depends on the distance $r$ between the object and the radar, as well as the speed of light $c$.

The IF signal then undergoes analog-to-digital conversion (ADC) and is subsequently sampled at a rate of $f_s$, producing $S$ samples per chirp. Assuming the radar is fitted with $M_r$ receive antennas, the number of samples per measurement will be $M_r \cdot S \cdot L$. These samples are represented as $\mathbf{R} \in \mathbb{C}^{M_r \times S \times L}$, and they constitute the fundamental information used to infer object-related information in RaDaR.

\paragraph{Range and velocity calculation} Suppose an object has a time-varying distance $r(t)=r_0+x(t)$, where $x(t)=vt$ is a function denotes the distance variation of an object moving at $v$ velocity. Thus, the round-trip time can be written as:
\begin{equation}
    \tau_{rt} = \frac{2r_0 + 2vt}{c}.
    \label{RT}
\end{equation}
After substituting (\ref{RT}) in (\ref{IF}) and performing some mathematical manipulations, the IF signal can be written as follows:
\begin{equation}
\small Y(t)=A_t A_r \exp \left(j\frac{4\pi}{c}\left(m v t^2 + m r_0 t + \frac{v}{\lambda_R} t + \frac{r_0}{\lambda_R}\right)\right),
\end{equation}
where $\lambda_R$ denotes the wavelength that corresponds to the operating frequency of the radar. Note that the first and last terms of the exponent have limited usefulness in extracting range and velocity information. The first term is very small, and the last term remains constant. In contrast, the second term provides valuable range information, while the third term enables the extraction of velocity information.

After sampling the IF signal with the ADC converter, a fast Fourier transform (FFT) is applied along the time samples direction, referred to as \textit{Range-FFT}, to determine the range of the object\footnote{Multiple ranges can result for an object. However, we consider the shortest one, which is usually produced by the nearest upper edge.}. The peak of the power spectral density output reveals the range information of the target object. Additionally, performing a second FFT on the chirp samples, known as \textit{Velocity-FFT}, helps determining the target object's velocity by observing the peak of the output spectrum.

\paragraph{Angle and direction estimation} The utilisation of MIMO antennas in radar systems enables the estimation of the angle of arrival (AoA). By implementing an additional FFT in the direction of the receive antenna samples, referred to as \textit{Angle-FFT}, the angular information can be extracted. Specifically, the variation in distance between the object and each receiving antenna causes a phase shift in the FFT peak, which corresponds to angular information. On the other hand, the object movement direction can be easily identified by observing changes in the AoA, or alternatively by checking which of the in-phase or quadrature components of the complex beat signal is leading in phase. Finally, performing range, velocity, and FFTs would result in \textit{radar cube}, which can be viewed as the stack of range-angle maps of each velocity value.

\begin{figure}
\centering
\includegraphics[scale=0.35]{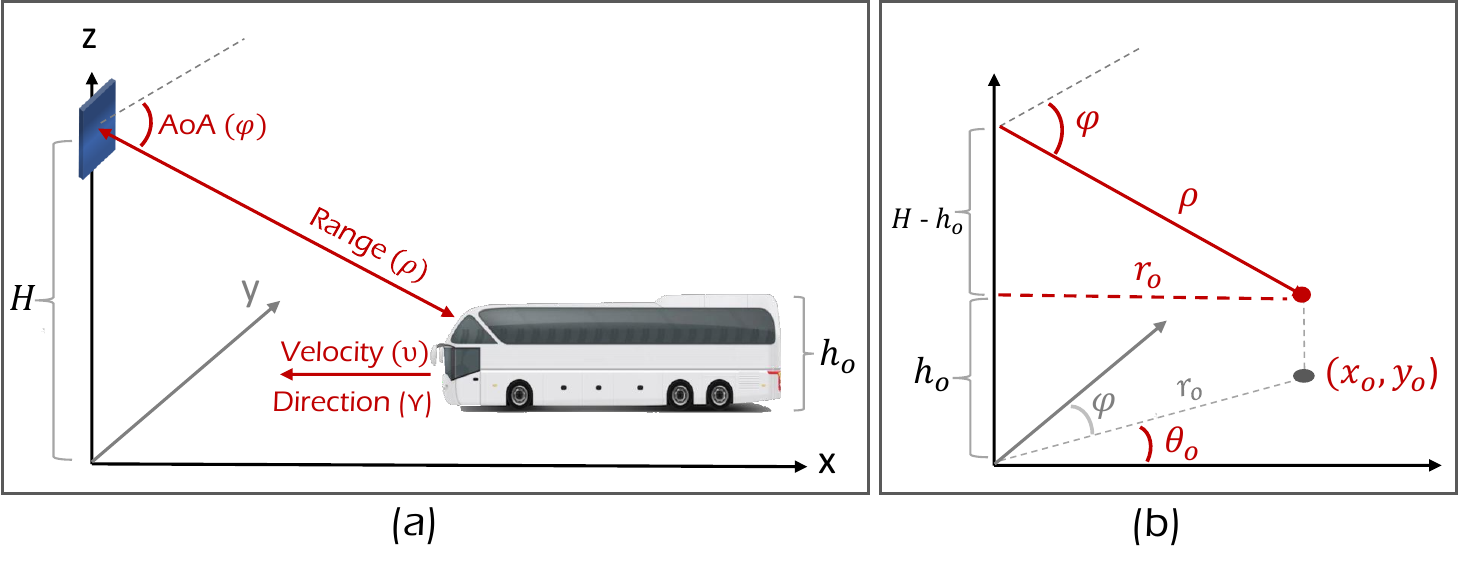}
\caption{(a) Radar is used to gather situational information, (b) this information is used to generate the localisation vector of the object. }
\label{RD}
\end{figure}

\section{Problem Description and Formulation}\label{PF}
Our main objective is to utilise radar measurements $\mathbf{R}$ for detecting objects and predicting forthcoming LoS beam blockages. This decision is facilitated by knowing the location information of the users and the potential obstacles. Once a stationary user $u$ is connected to the network, the SBS activates the radar sensor to monitor the surrounding area. Assuming that a number of objects denoted as $O$ are detected in the $i$th measurement $\mathbf{R}_i$. Each object $o \in O$ will be continuously monitored until it exits the radar's field of view (FoV). During this monitoring process, the object's situational parameters, which include its range $\rho$, velocity $\upsilon$, AoA $\varphi$, and direction $\curlyvee$ described in Fig. \ref{RD}(a) are extracted. Based on the height of the radar $H$ and the height of the object $h_o$, which will be discussed in Section \ref{key}, we utilise the situational information to construct a six-dimensional localisation vector $\mathcal{L}$ for the object, given by $\mathcal{L}_o=[r_o,x_o,y_o,\theta_o,v_o,n_o]$, where: 
\begin{equation}
\begin{aligned}
& r_o=\sqrt{ \left(\rho^2-(H-h_o)^2\right)} \\
& x_o=r_o \sin (\varphi) \\
& y_o=r_o \cos (\varphi) \\
& \theta_o=\tan ^{-1} y_o / x_o\\
& v_o=\upsilon, \hspace{0.1cm} n_o=\curlyvee.
\end{aligned}
\end{equation}
These features are derived by placing the SBS at the Cartesian origin and considering the x-y plane. Specifically, we define $r_o$ as the distance from the SBS to the object, and $x_o$ and $y_o$ as the object’s x and y coordinates, respectively. Additionally, we use $\theta_o$ to represent the angle between the positive $x$-axis and the line passing through ($x_o$,$y_o$) and the origin. The object's speed is denoted by $v_o$, while $n_o$ denotes its direction. Please refer to Fig. \ref{RD}(b) for further illustration.

Given a set of stationary users $U$, we assign to each user $u \in U$ a localisation vector denoted as $\mathcal{L}_u=[r_u,x_u,y_u,\theta_u]$. These features are similar in their definition to those related to the detected object and are obtained by converting the user’s GPS information to Cartesian using the universal transverse Mercator tool \cite{[1.34]}. Then, for each user $u$, we generate $S_{u,o}=\{\mathcal{L}_{u},\mathcal{L}_{o}\}$, with the objective of classifying whether this data sample results in a future blockage $b\in \{0,1\}$. 0,1 denote beam non-blockage or blockage, respectively.  Moreover, we estimate the remaining time until the obstacle obstructs the LoS connection, denoted as $T_b$ and is defined as:
\begin{equation}
T_{b}=\left\{\begin{array}{l}
\hspace{0.1cm} \xi\hspace{0.1cm}, b=1, \hspace{0.2cm} \forall \xi \in \mathbb{R}^{+} \\
\hspace{-0.1cm} -1, b=0,
\end{array}\right.
\end{equation}
where $-1$ indicates that the value is not applicable due to the absence of potential blockage. Therefore, $s_{u,o}=\{b_{u,o}, T_{b_{u,0}}\}$ is defined as the labels associated with each data sample $S_{u,o}$.

The user/object localisation information could be leveraged to intelligently handle channel disruptions and enhance network reliability. To formulate that, our objective is to design an ML model, denoted as $\Psi_\Theta(S)$, that can simultaneously perform classification and regression. This model takes in user-object data samples $S$ and generates prediction $\hat{s}$. These predictions are governed by a set of parameters $\Theta$ adapted based on a labeled dataset $D=\{S_{u,o}, s_{u,o}\}$. The model aims to maximise the probability of accurately predicting link disconnections while minimising the error associated with predicting the blockage remaining time, as given below:
\begin{equation}
\max _{\Psi_\Theta(S)} \prod_{u=1}^{U} \mathbb{P}\left(\hat{b}_{u,o}=b_{u,o} \mid S_{u,o}\right), \hspace{0.2cm}\forall u \in U,\hspace{0.1cm} \forall o \in O
\end{equation}

\begin{equation}
\min _{\Psi_\Theta(S)} \sum_{u=1}^{U} (|\hat{T}_{b_{u,o}}- T_{b_{u,o}}|),\hspace{0.2cm} \forall u \in U, \hspace{0.1cm} \forall o \in O
\end{equation}

\section{The Proposed RaDaR Framework} \label{Framework}
In this section, we present RaDaR, an ML-based approach for predicting beam blockages in beyond fifth-generation (B5G) and sixth-generation (6G) networks using radar data. RaDaR utilises a dual-output NN model trained on bundles of radar measurements to provide the system with real-time blockage handling intelligence, thereby improving the performance of next-generation networks.  In the next subsections, we provide an in-depth explanation of the proposed solution.

\subsection{Overview and Schematic Diagram}\label{key}
\begin{figure}
\centering
\includegraphics[scale=0.52]{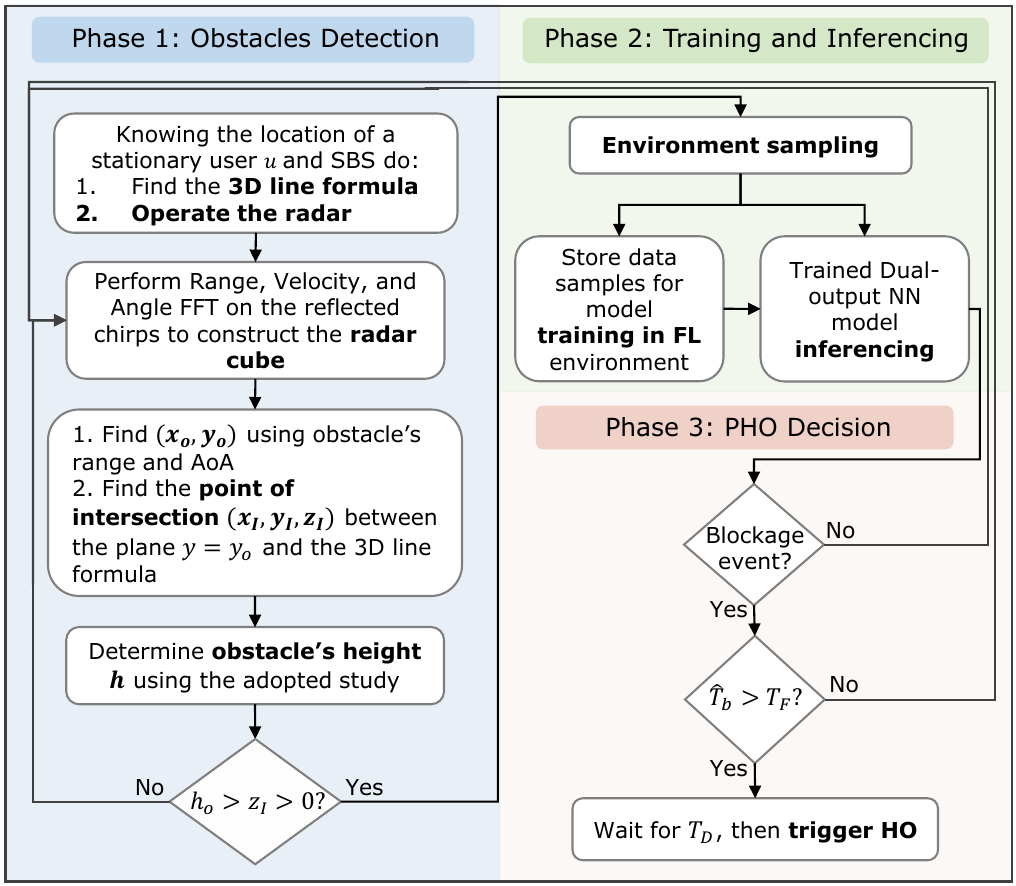}
\caption{Schematic diagram of the proposed framework.}
\label{SD}
\end{figure}

This study focuses on practical communication systems, particularly in the context of ultra-dense networks (UDNs). UDNs densify SBSs and LoS links per unit area. However, downscaling communication systems have complicated the challenge of mobility, particularly in dynamic areas like smart cities. SBSs use narrow directive beams to connect users to the network, making the UE’s position relative to the SBS critical for service continuity. However, the presence of mobile objects can obstruct LoS links, leading to fluctuations in data rates. Conventional wireless networks can only detect the presence of blockages when the user's throughput fluctuates or when the link is disconnected. This concludes that the network is reactive to blockage events, resulting in poor performance. To overcome this challenge, wireless networks must shift from reactiveness to proactiveness by having a sense of their surroundings. Proactiveness must be integrated as a key dimension using existing sensing modalities to improve wireless networks. Therefore, we adopt radar sensors where each SBS is equipped with an FMCW radar to monitor the coverage area. The information obtained from radars is vital in dealing with link blockages and controlling the communication system. Fig. \ref{SD} illustrates the proposed framework’s schematic diagram, which consists of three main phases: obstacles detection phase, training and inferencing phase, and PHO decision phase.

\subsubsection{The obstacles detection phase} 
\begin{figure*}
\centering
\includegraphics[scale=0.25]{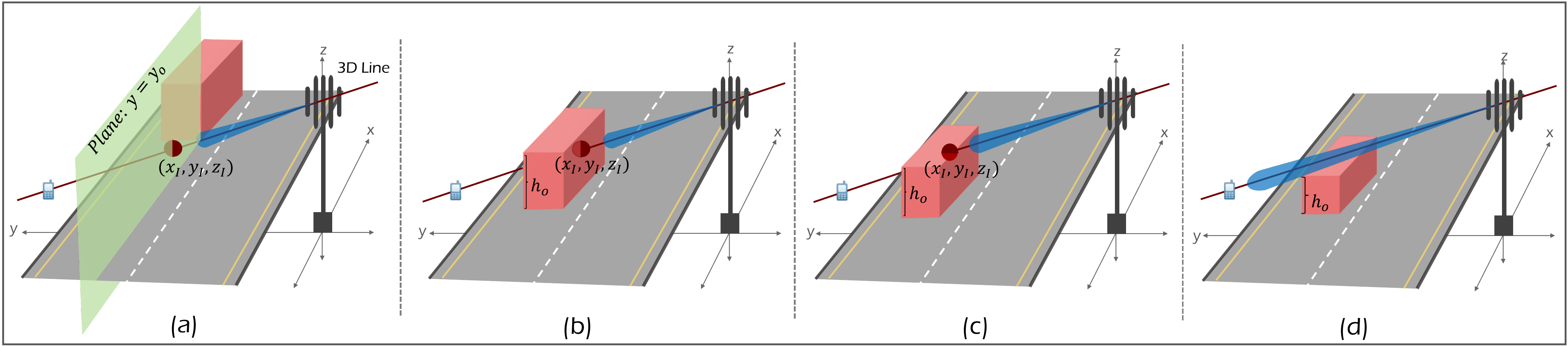}
\caption{Obstacle detection analysis: (a) the use of 3D line equation and $y=y_o$ plane to determine the point of intersection. Assuming object's height is $h_o$, then (b) $h_o > z_I$, means a blockage, (c) $h_o = z_I$, means a blockage (d) $h_o < z_I$, means no blockage.}
\label{Int}
\end{figure*}

Not every object detected by the radar will necessarily act as a blockage disrupting the communication channel. The occurrence of a blockage event in our system model is highly dependent on the position of the antenna array, the UE, and objects, as well as the geometry of the objects, particularly their height. Therefore, it is imperative to accurately identify actual obstacles to prevent unnecessary HOs. To achieve this, when a stationary user is connected to the network, the serving SBS uses the three-dimensional (3D) location information of the user ($x_u,y_u,z_u$) and the antenna array ($x_A,y_A,z_A$) to determine the formula of a 3D line crossing them. This equation can be expressed as a vector equation in mathematical terms as follows:
\begin{equation}
\begin{aligned}
& <x, y, z>=<x_A, y_A, z_A>+\eta<\iota, \varsigma, \kappa>\\
& (\iota, \varsigma, \kappa)=(x_u-x_A, y_u-y_A, z_u-z_A)\\
\end{aligned}
\end{equation}
where $\eta$ is a parameter describing a particular point on the line, and $<\iota, \varsigma, \kappa>$ is the direction vector. At the same time, the SBS activates the radar. Upon detecting an object, the framework generates its localisation vector. Subsequently, the framework generates the plane $y=y_o$ and determines the point of intersection between the 3D line equation and the plane, denoted as $(x_I,y_I,z_I)$, as illustrated in Fig. \ref{Int}(a).

To determine the height of a detected object, we employ one of the several studies available in the literature that tackled this issue through radar technology \cite{[1.22],[1.23]}. The most appropriate technique involves using radar fingerprints and residual networks (ResNets) to classify objects in real time. Then, this technique determines the objects' heights by referencing a predefined table that includes the dimensions of each classified object \cite{[1.23]}. Therefore, we assume that our framework adopts this approach to determine the height of the detected object. By knowing the point of intersection and the height of the object, the framework compares $z_I$ with the $h_o$ to determine if the object can potentially obstruct the LoS link. If $h_o$ is greater than or equal to $z_I$, this means the object will block the connection, as demonstrated in Fig. \ref{Int}(b) and \ref{Int}(c). Conversely, if $h_o$ is less than $z_I$, this means the object most likely will not block the connection, as shown in Fig. \ref{Int}(d). 

\subsubsection{Training and inferencing phase} The next phase involves sampling the surrounding environment and collecting the necessary data samples for supervised model training. The framework acquires the requisite dataset to train the NN model for predicting the status and time of blockages by localising users and objects. The training process is performed offline using the FL algorithm, in which each SBS contributes its collected datasets to collaborate on model training. Afterward, the trained model is deployed online to perform inferencing.

\subsubsection{PHO decision phase} If a blocking event is anticipated, the primary objective of the framework is to prevent user shadowing and maintain the connection by proactively deciding to perform a HO. This decision is supported by predicting the remaining time until the obstacle blocks the LoS links $\hat T_b$. Knowing this time enables better planning for the optimal time to perform HO and maintain the QoE at its highest possible level. However, it is important to measure the total time required by the proposed framework, denoted as $T_F$, which comprises three main sub-time parameters as follows:
\begin{equation}
T_F = T_R + T_{Inf} + T_{HO},
\label{TF}
\end{equation}
where $T_R$ is the time duration starting from activating the radar until processing the radar cube information for a single measurement. $T_{Inf}$ represents the model inferencing time, and $T_{HO}$ indicates the time required to switch the user to another stable connection. Moreover, picking the optimal time instant for initiating proactive HO is facilitated by introducing a new time parameter called the delay time ($T_D$). This parameter represents the idle time between the completion of inferencing and the triggering of HO, and is defined as follows:
\begin{equation}
T_D \leq \hat T_b - T_F.
\label{TD}
\end{equation}

\subsection{Radar Measurements Processing}
The $T_R$ parameter entails several sub-processes that the framework executes to prepare for the next stage of generating the data samples. Table \ref{RP} presents the typical radar system parameters considered in this study \cite{[1.19]}. Initially, we measure the duration of each measurement, denoted as $T_m$, which involves transmitting 128 chirps, each lasting for 60 $\mu s$, requiring a total of 8.3 ms. Next, we calculate the maximum time a radar signal can remain in the air, given that the maximum radar range is set to 100 meters. This time is expressed in microseconds; however, it can be disregarded since the framework’s timescale is in the millisecond range. Then, we measure the sampling time $T_s$ required to perform ADC and sampling of the received reflected signals. $T_s$ is calculated by dividing the number of samples per measurement by the sampling rate and is found to be 26.2 ms.

\begin{table}
\caption{Radar system parameters \cite{[1.19]}.}
\centering
\fontsize{8.5pt}{8.5pt}\selectfont
\begin{tabular}{l c}
\hline \\[-0.7em]
\textbf{Parameter}                & \textbf{Value} \\ \hline
\hline \\[-0.8em]
No. of transmitters               & 1              \\ \\[-0.9em]
No. of receivers                  & 4              \\  \\[-0.9em]
No. of chirps                     & 128            \\  \\[-0.9em]
Start frequency ($f_c$)           & 77 GHz           \\  \\[-0.9em]
Chirp slope ($m$)                & 15015 GHz/s      \\ \\[-0.9em]
Chirp duration ($\tau_c$)             & 60 $\mu s$             \\  \\[-0.9em]
Chirps pause time ($\tau_p$) & 5 $\mu s$             \\  \\[-0.9em]
No. of samples per chirp          & 256            \\  \\[-0.9em]
Sampling rate                     & 5 MHz            \\  \\[-0.9em]
Max range                    & 100 m           \\ \hline
\end{tabular}
\label{RP}
\end{table}

After acquiring the radar measurements in the form of samples, the next step is to measure the time taken to perform FFT and generate the radar cube, represented as $T_{FFT}$. Assuming an FFT process has a complexity of $\mathcal{O}(n\log n)$, and a single token $n$ requires one nanosecond to execute, we estimate that generating a radar cube per measurement requires performing three FFTs, resulting in a processing time of 6 ms \cite{[1.24]}. Furthermore, we identify the time required to classify the detected object, denoted as $T_c$, in the adopted work. To ensure real-time object classification, we refer to a previous study \cite{[1.25]} that demonstrates the ResNet-50 model's inference time to be 26 ms, which is considered near real-time. Therefore, we assume that the adopted work requires a similar time to classify the detected object. In summary, the following equation defines the sub-processes times covered under $T_R$:
\begin{equation}
    T_R=T_m+T_s+T_{FFT}+T_c.
\end{equation}
Based on the above assumptions and discussion, $T_R$ is calculated to be around 66.5 ms for our illustrative realistic scenario.

\subsection{Federated Learning Design for Model Training}
This study aims to select a model that meets various requirements, including high prediction accuracy, low inference latency, collaborative training capability, and simultaneous classification and regression ability. Through careful investigation, we design a three-hidden layer NN model that processes user-object data samples and produces dual predictions of blockage status and time, as depicted in Fig. \ref{NN}. Furthermore, our study aligns with the current research trend of using the FL approach instead of the centralised learning mechanism to provide several benefits, such as safeguarding data privacy, improving bandwidth efficiency, and promoting scalability and knowledge sharing. This approach enables the framework to be generalised by learning from different scenarios, facilitating its deployment in new development areas.

\subsubsection{Offline learning phase}
\begin{figure}
\centering
\includegraphics[scale=0.45]{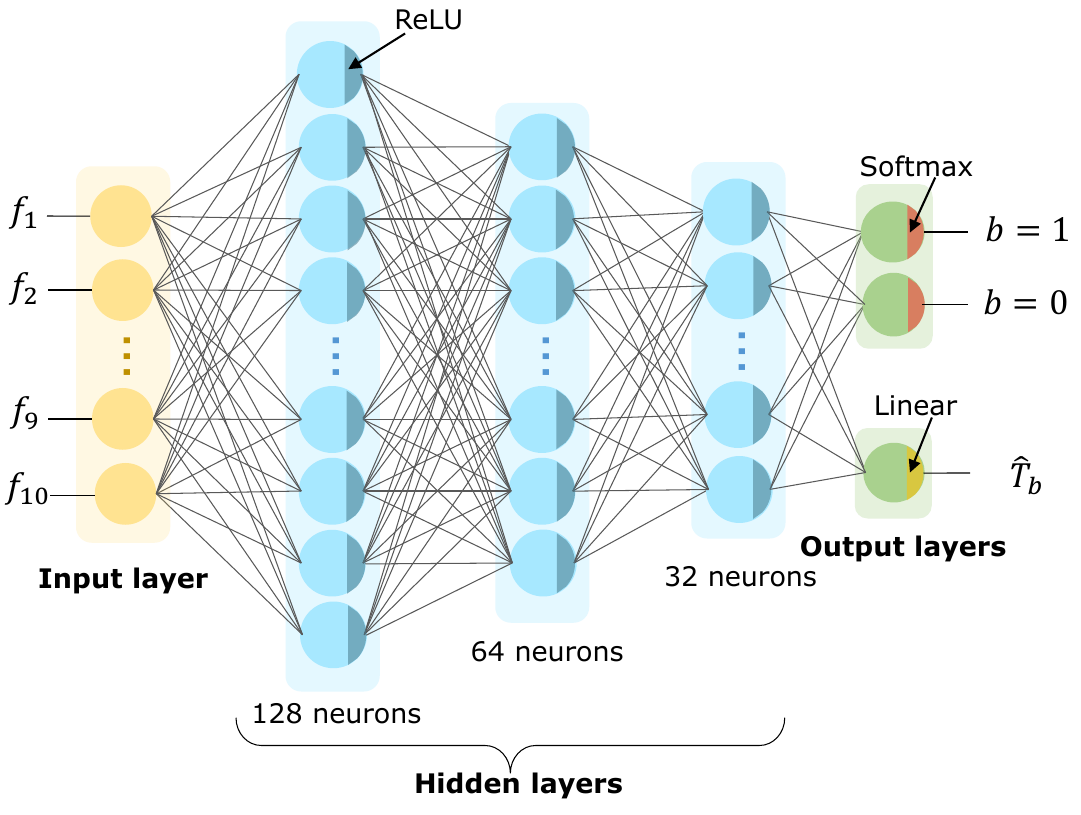}
\caption{The structure of the developed dual-output NN model.}
\label{NN}
\end{figure}

The proposed dual-output NN model is trained offline using the FL mechanism. The model comprises an input layer that receives ten features \{$f_{1}$, $f_{2}$ ,..., $f_{10}$\} of user-object data samples, followed by three hidden layers with 128, 64, 32 neurons, respectively. The model has two output layers: a classification layer with two neurons activated by the softmax function and a regression layer with a single neuron activated by the linear function. The model’s architecture is depicted in Fig. \ref{NN}. The loss functions used are mean absolute error (MAE) and sparse categorical crossentropy. Additionally, the model’s optimiser, learning rate, batch size, and epochs are set to Nadam, 0.001, 100, 10, respectively. The training process involves SBSs acting as clients participating in model training. In this study, we use five clients, but the framework is scalable and can accommodate more clients. Each SBS utilises its user-object data samples to iteratively train the NN model and reports the model updates to a central server located, for example, in a macro-BS or the cloud. Moreover, the server follows the federated averaging method \cite{[1.26]} to aggregate the shared model parameters by computing their weighted average. The weight of each update is proportional to the number of data points of each client. In addition, the number of FL communication rounds is controlled using the delta-based FL stopping technique \cite{[1.27]} to avoid suboptimal or unnecessary communication rounds.

\paragraph{{Development dataset}} The NN model is trained and evaluated by exploiting the real-world data from scenario 30 of the DeepSense 6G testbed \cite{[1.19]}. This testbed closely resembles our system model, wherein a transmitter and receiver located on opposite sides of a two-way city street. The scenario offers diverse data modalities, including radar measurements and blockage information with associated timestamps. By computing the difference between timestamps, we can label user-object samples and determine when the detected object will obstruct the LoS link. However, in scenario 30, every object is blocking the LoS link due to the low height of the BS, which does not necessarily reflect practical network deployments. In contrast, our system considers more practical communication systems, where the detected objects may or may not cause blockages depending on the height of the SBS. Additionally, scenario 30 only involves a single BS communicating with a single stationary user, whereas our study targets wireless networks that involve multiple SBSs and users. To address these limitations, we analyse and augment scenario 30 by adjusting various parameters of the testbed, such as street and object dimensions, radar height, user position, dataset size, and blockage samples ratio. By using different distributions, we aim to reflect more practical wireless networks and generate multiple distinct environments. Table \ref{Augmented} summarises the parameters that are modified to generate new non-IID datasets.

\begin{table*}
\caption{The parameters that have been adjusted in the testbed of scenario 30 to reflect a practical wireless communication system. [a,b] indicates the range in which values are chosen based on the corresponding distribution.}
\fontsize{8.5pt}{8.5pt}\selectfont
\centering
\begin{tabular}{|c c c c c c c c c c|}
\hline\\[-0.7em]
\textbf{SBS} &
  \textbf{Distribution} &
  \textbf{Street [m]} &
  \begin{tabular}[c]{@{}c@{}}\textbf{Object}\\  \textbf{height [m]}\end{tabular} &
  \begin{tabular}[c]{@{}c@{}}\textbf{Radar}\\ \textbf{height [m]}\end{tabular} &
  \begin{tabular}[c]{@{}c@{}}\textbf{User\_y} \\ \textbf{[m]}\end{tabular} &
  \textbf{Object\_y [m]} &
  \textbf{Speed [mps]} &
  \begin{tabular}[c]{@{}c@{}}\textbf{No. of}\\ \textbf{samples}\end{tabular} &
  \textbf{Block: Non-block} \\ \hline\\[-0.8em]
SBS1 & Uniform  & [-20, 20] & [1, 4.5] & 3 & 12 & [1, 11] & [3, 9]  & 10,000 & 10\% : 90\% \\ \hline \\[-0.8em]
SBS2 & Gaussian & [-30, 30] & [1, 4.5] & 4 & 13 & [1, 12] & [3, 11] & 15,000 & 25\% : 75\% \\ \hline \\[-0.8em]
SBS3 & Gamma    & [-40, 40] & [1, 4.5] & 5 & 14 & [1, 13] & [3, 13] & 30,000 & 50\% : 50\% \\ \hline \\[-0.8em]
SBS4 & Binomial & [-50, 50] & [1, 4.5] & 6 & 15 & [1, 14] & [3, 15] & 25,000 & 75\% : 25\% \\ \hline \\[-0.8em]
SBS5 & Poisson  & [-60, 60] & [1, 4.5] & 7 & 16 & [1, 15] & [3, 17] & 20,000 & 90\% : 10\% \\ \hline \\[-0.8em]
SBS6 & Beta     & [-50, 50] & [1, 4.5] & 5 & 13 & [1, 12] & [3, 9]  & 2,000  & 50\%: 50\%  \\ \hline
\end{tabular}
\label{Augmented}
\end{table*}

\begin{figure}
\centering
\includegraphics[scale=0.4]{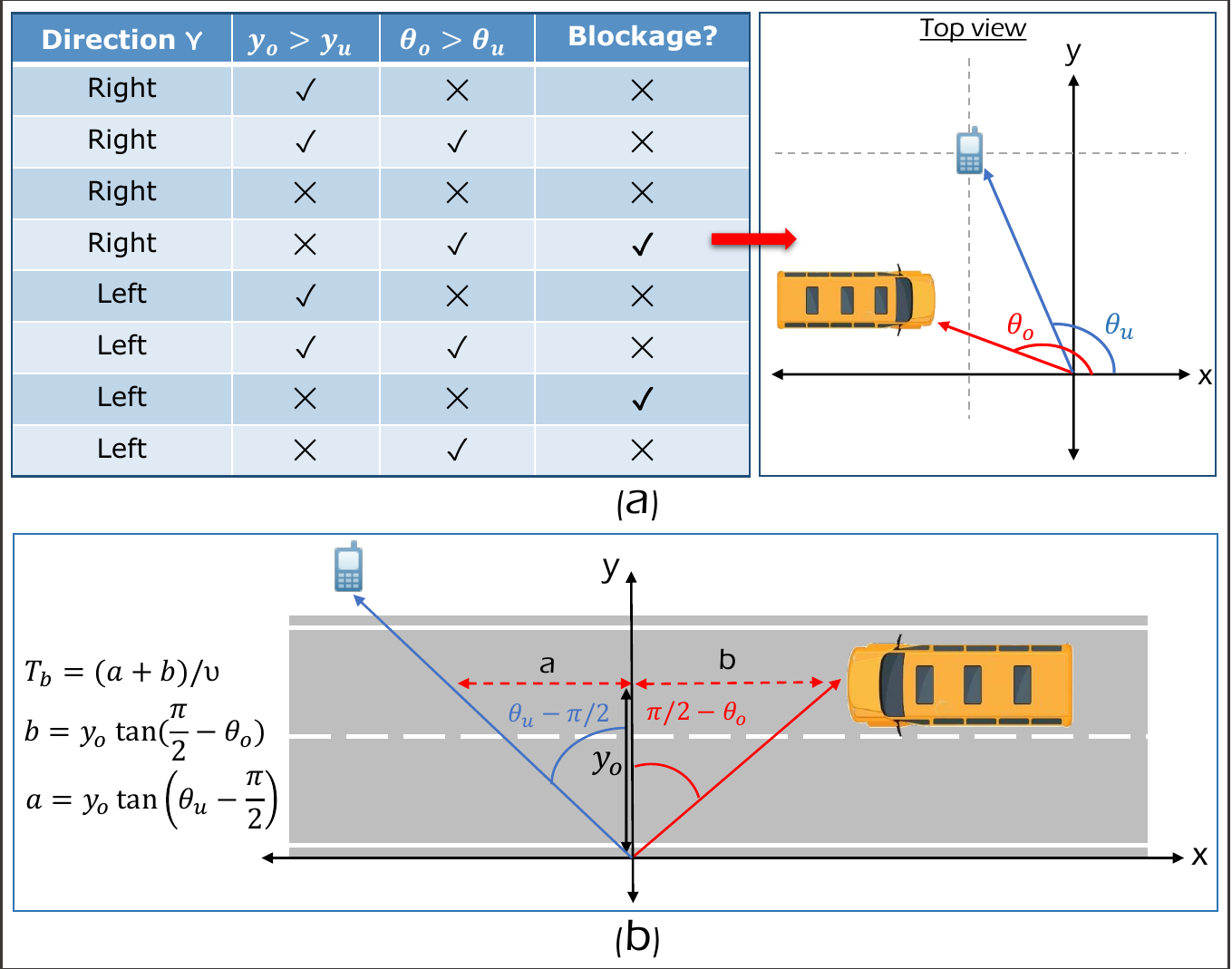}
\caption{Labeling user-object data samples: (a) blockage status of various cases, (b) A case demonstrating the calculation of the $T_b$.}
\label{Labels}
\end{figure}

Through these modifications and augmentations, we now have a more realistic communication system with six SBSs, each having its own user-object data samples. If $h_o > z_I$, we use $[\curlyvee, y_o, y_u, \theta_o, \theta_u]$ as a set of features to judge blockage status. The table in Fig. \ref{Labels}(a) shows all possible cases, from which we can observe that only two cases will lead to blockages. Fig. \ref{Labels}(a) also depicts one of the blockage cases. Similarly, to determine the $T_b$, the position information, movement direction, and speed are important. Using trigonometry, $T_b$ can be determined, as shown in Fig. \ref{Labels}(b).

\begin{figure}
\centering
\includegraphics[scale=0.6]{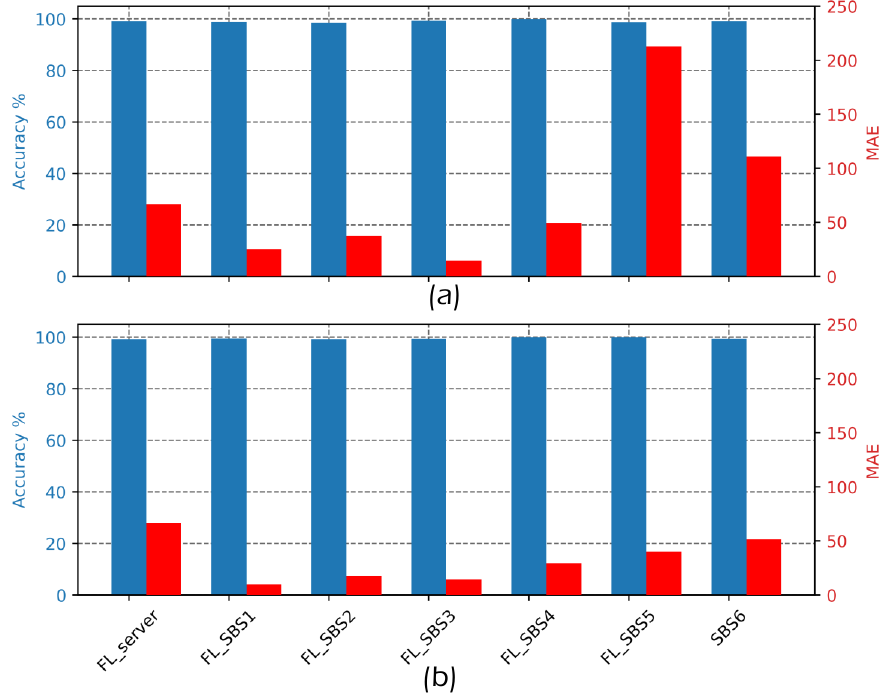}
\caption{The classification and regression performance of the dual-output NN model (a) without and (b) with tuning.}
\label{MPerf}
\end{figure}

\paragraph{{Model evaluation}} After preparing the data samples with their respective labels for each SBS, 2000 samples were allocated for model evaluation, while the remainder were reserved for training. From the evaluation dataset, we randomly select a small dataset to form a total of 5000 samples used for evaluating the model's performance at the server. Fig. \ref{MPerf}(a) displays the classification and regression performance of the dual-output NN model at the SBSs and the server after 30 communication rounds of training using FL. While the classification accuracy was near optimal for all SBSs, there were variations in the MAE values, which is attributed to the non-IID properties of the training datasets, where each dataset was obtained from a different distribution representing a distinct environment. To further enhance the performance, each SBS was permitted to personalise the model by utilising 500 data samples and performing a few rounds of tuning. Fig. \ref{MPerf}(b) illustrates the final performance of the trained models, depicting a further enhancement in the prediction accuracy of the model for each SBS as a result of personalisation.

\paragraph{{Knowledge transfer case study}} During the augmentation process, we have created an additional environment, named SBS6, representing a new deployment area with few data samples, as outlined in Table \ref{Augmented}. The purpose is to assess the generalisation and scalability of the proposed framework for supporting knowledge transfer and the rapid deployment of new SBSs. SBS6 did not participate in the FL training process, and the server pushed the latest version of the trained NN model to this SBS to initiate its operation. The performance of SBS6 is presented in Fig. \ref{MPerf}, showing a remarkably high classification accuracy. However, the regression accuracy is inferior to that of the first four SBSs, although it is better than SBS5, which participated in the training process. After tuning, the regression performance is significantly improved by a percent of 53\%. These results demonstrate that the proposed framework is scalable and can support the rapid deployment and operation of new network sites.

\subsubsection{Online inference phase}
After training and personalising the models, each SBS is ready to use its personalised model to predict the occurrence of blockages and, hence, the stability of the LoS beams. When a stationary user is connected to the network, the SBS creates samples of user-object data and feeds them to the model for inference. The model then predicts the status of the blockages and the time until the object blocks the link. Given that the dual-output NN model is much simpler than ResNet-50, the $T_{Inf}$ should be much lower than the 26 ms inference time of ResNet-50. We measured the $T_{Inf}$ using a standard personal computer and found it to be around 1 ms. Although it should be even lower, for the sake of analysis, we assume $T_{Inf}$ equals 1 ms.

\subsection{Optimal HO Trigger Point }
After detecting an obstacle and determining the blockage time, the framework must notify the network to perform a HO and ensure user connectivity. The crucial question is, what is the best time instant or distance point to trigger the HO process and switch the user to alternative stable links? Defining the $T_D$ parameter in equation (\ref{TD}) should help in finding such instant/point by introducing the following formula:
\begin{equation}
\Lambda \leq v \times T_D
\end{equation}
where $\Lambda$ represents the points at which the network can trigger PHO and maintain stable connections for the user each time an obstacle crosses that points. Since our framework is QoE-aware, it aims to delay the HO process until reaching the optimal point ($\Lambda _{opt}$), which corresponds to the maximum tolerable $T_D$ and is given by:
\begin{equation}
\Lambda_{opt} = v \times T_D^{max}.
\end{equation}
By adopting this approach, our framework ensures a seamless user experience while avoiding obstacle's disruption.

\subsection{PHO Procedure and Latency Minimisation}\label{PHOLatency}
In conventional wireless networks, when there is a degradation in the signal quality of a user's connection, a HO mechanism is initiated based on predetermined events detected through measurement reports \cite{[1.29]}. If beamforming technology is being utilised and the LoS beam is disconnected, then several steps must be taken to re-establish a stable connection. The steps are beam failure detection, beam failure recovery, cell search, and contention-based or free random access. Each of these steps requires execution time that combined would result in high latency $\sim$ 312.2 ms, affecting the reliability of the communication system.

This study introduces a new HO event, called a block event, which is defined by the detection of an obstacle that could potentially block the LoS user connection. If a block event is detected, the network should take preemptive measures to prevent channel interruption. Our proposed framework is proactive in nature, eliminating the first two steps of the HO process and performing the cell search step in advance while the user is still connected. Therefore, the HO latency boils down to the latency associated with experiencing either contention-based or free random access. This framework considers the worst-case of contention-based random access that requires about 80 ms \cite{[1.16]}. As a result, the last parameter of equation (\ref{TF}), i.e., $T_{HO}$ equals to 80 ms. Now that we have determined all the values of the parameters in equation (\ref{TF}), which total of 147.5 ms represents the execution time of our proposed framework. Finally, it is important to note that the performance of RaDaR framework depends heavily on the hardware specifications of the radar, server, and network. As hardware specifications improve, the framework's execution time will reduce, resulting in a further enhanced framework.

\section{Performance Evaluation and Results}\label{PEval}
This section examines the efficacy of the proposed RaDaR framework in improving the operation of high-frequency communication systems, such as UDNs, by effectively and preemptively predicting the occurrence of beam blockages and implementing appropriate measures to guarantee uninterrupted connectivity for users while preserving high levels of QoE.

\subsection{Dual-Output Model Development}
After training the dual-output NN model in an FL environment and performing personalisation, the resulting models are now ready to be tested in practical scenarios. To further evaluate the models, we define a new parameter called the PHO success rate ($\mathfrak{S}_{PHO}$), which represents the percentage of successfully detecting beam blockages and performing PHO. $\mathfrak{S}_{PHO}$ is calculated by dividing the number of samples successfully performing PHO by the total number of samples. Using the same evaluation dataset of the SBSs, the $\mathfrak{S}_{PHO}$ results for each SBS are presented in the 0\% column of Table \ref{SRPS}. The figures reveal variations in the results, where the $\mathfrak{S}_{PHO}$ of some SBSs is unsatisfactory, while others exhibit better performance. After careful investigation, we discovered that the inconsistent behaviour is due to the inaccurate prediction of the blockage time $\hat T_b$. Occasionally, the models predict $\hat T_b$ to be greater than the actual time, which causes the user to get blocked before completing the PHO process because the RaDaR framework is QoE-aware and waits for the maximum $T_D$ to reach the optimal point before performing PHO.

To address this issue, we have introduced a new parameter named percent shift ($\mathfrak{P}_{Shift}$), which aims to mitigate the impact of blockage by reducing the predicted blockage time by a certain percentage. This may result in slightly reduced QoE for the user due to earlier PHO. However, we believe that maintaining a stable connection with slightly reduced QoE is preferable to losing the connection and having to reestablish it. Our experiments show that the introduction of $\mathfrak{P}_{Shift}$ has significantly improved the $\mathfrak{S}_{PHO}$ for all SBSs, as shown in Table \ref{SRPS}. Nevertheless, there exists a trade-off between selecting the values of $\mathfrak{P}_{Shift}$ and the perceived QoE. It is imperative to note that higher values of $\mathfrak{P}_{Shift}$ can yield further improvements. However, such improvements may necessitate even earlier PHOs, thereby potentially degrading the user's QoE. Hence, we have selected the value of 10\% for SBS1, 2, 3, and 6 and the value of 8\% for SBS4 and SBS5.

\begin{table}
\caption{Study of $\mathfrak{S}_{PHO}$ [\%] versus $\mathfrak{P}_{Shift}$ for different SBSs.}
\fontsize{8.5pt}{8.5pt}\selectfont
\centering
\begin{tabular}{|c|c|c|c|c|c|c|}
\hline \\[-0.8em]
\backslashbox{\textbf{SBS}}{$\mathbf{\mathfrak{P}_{Shift}}$} &
  0 \% &
 2 \% &
  4 \% &
  6 \% &
  8 \% &
  10 \% \\ \hline \\[-0.8em]
SBS1 & 55.4 & 69.6          & 80.1 & 87.2 & 91.4          & \textbf{93}   \\ \hline \\[-0.8em]
SBS2 & 36.9 & 66.5          & 81.5 & 87.4 & 89.8          & \textbf{90.6} \\ \hline \\[-0.8em]
SBS3 & 70   & 82.4          & 85.6 & 86.7 & 87.7          & \textbf{88.2} \\ \hline \\[-0.8em]
SBS4 & 96.9 & 99.1 & 99.1 & 99.1 & \textbf{99.3}          & 99.3          \\ \hline \\[-0.8em]
SBS5 & 41.1 & 87.9          & 96.6 & 98.2 & \textbf{98.8} & 98.8          \\ \hline \\[-0.8em]
SBS6 & 30.5 & 62.3          & 85.3 & 93.2 & 95.7          & \textbf{97.3} \\ \hline
\end{tabular}
\label{SRPS}
\end{table}

\begin{figure}
\centering
\includegraphics[scale=0.32]{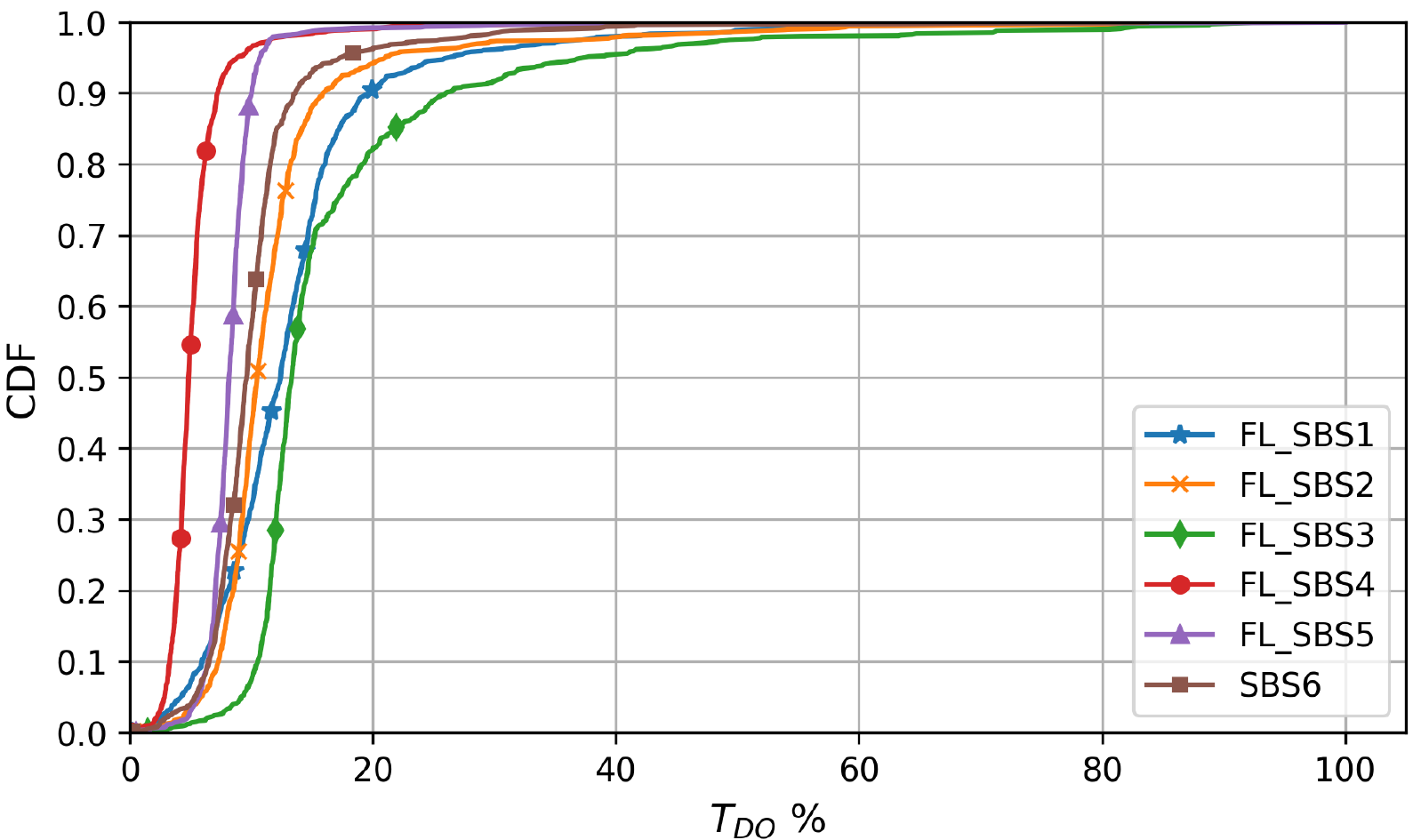}
\caption{The distribution of the $T_{DO}$ samples that lead to a successful PHO for different SBSs.}
\label{TDO}
\end{figure}

To investigate the effect of the $\mathfrak{P}_{Shift}$ parameter, Fig. \ref{TDO} plots the cumulative distribution function (CDF) for every sample $i \in D$ results in successful PHO, with respect to time delay offset $T_{DO}$. Here, $T_{DO}$ is defined as:
\begin{equation}
    T_{DO_{i}}=\frac{T_{D_{i}}^{max} - \hat T_{D_{i}}}{T_{D_{i}}^{max}} \times 100\%, \hspace{0.1cm} \forall \hat T_{D_{i}} \leq T_{D_{i}}^{max}, i \in D
\end{equation}
where $T_{D}^{max}$ is the maximum actual delay time before triggering PHO, and $\hat T_{D}$ is the predicted time delay given as $\hat T_{D}=\hat T_{b}-T_{F}$. $T_{DO}$ indicates how far the $\hat T_{D}$ from the actual one; the closer the values to zero, the better the performance. Fig. \ref{TDO} illustrates the overall performance of the SBSs in predicting blockages and successfully executing PHO, with only small variations in the values of the $T_{DO}$. All SBSs have more than 80\% of their samples with $T_{DO}$ is less than 20\%, which highlights the superiority of the proposed framework in proactively predicting blockages and performing PHO at the time/point that maintains the user’s QoE as high as possible. Thus, the framework of each SBS is now ready for deployment and further investigation.

\subsection{Experimental and Simulation Setup}
We employ a hybrid approach that integrates real-world and simulated environments to assess the overall performance of the proposed RaDaR framework. The simulated environments are derived from scenario 30 to reflect more practical wireless networks and generate multiple distinct environments. The simulation experiments are implemented based on Python installed on a Windows operating system with an Intel Xeon CPU E5-2620 @ 2GHz and 16GB RAM. The key performance metrics are the RSS, network latency, and throughput.

\subsection{Performance Analysis}
\begin{figure}
\centering
\includegraphics[scale=0.45]{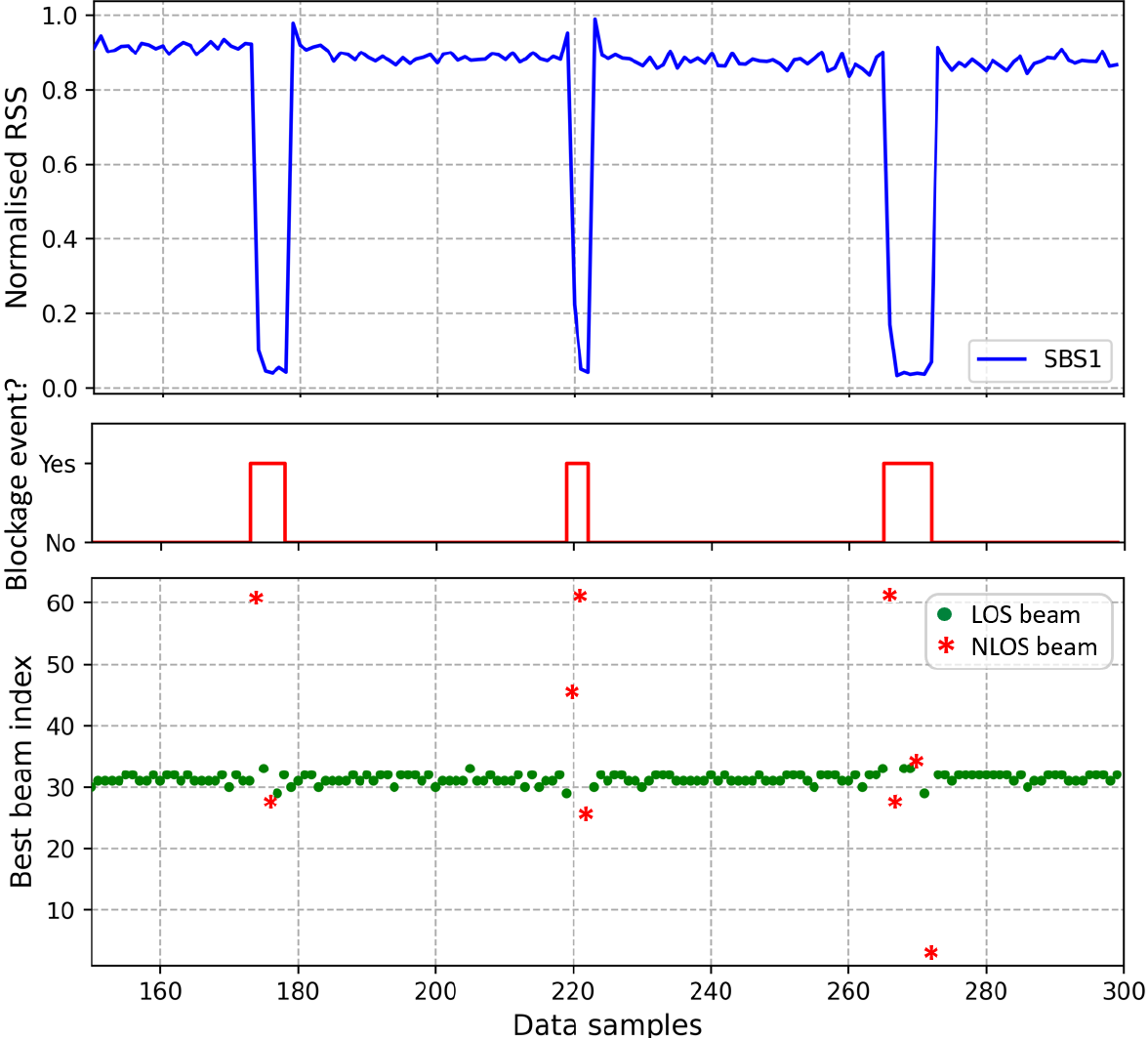}
\caption{The normalised RSS, blockage events, and best beam index in a Reactive-HO communication system.}
\label{RHO}
\end{figure}

This section reveals the effectiveness of the proposed RaDaR framework in comparison to a conventional wireless network that lacks proactive blockage prediction techniques and only responds reactively to blockage events, which we refer to as Reactive-HO. Initially, we analyse the effect of blockages on the RSS at the user’s end in Reactive-HO communication systems, using the diverse information available in the DeepSense scenario 30 testbed. Fig. \ref{RHO} illustrates the normalised RSS of a stationary user, blockage events, and the best beam index at each data sample. The results demonstrate unstable performance, as the RSS deteriorates each time an obstacle obstructs the LoS beam serving the user. This performance is unsuitable for time-sensitive services and data-intensive applications, such as high-definition video streaming, which require fast and reliable wireless connections to guarantee an uninterrupted and smooth user experience. Furthermore, the figure demonstrates that in the absence of obstacles, the best-serving beam is consistently limited to a few fixed beams. However, when the communication channel is blocked, the best beam index varies and can be any beam from the beamforming codebook, determined by the beam that gives the highest received power after reflection from the environment. These findings highlight the importance of proactive blockage prediction techniques in achieving reliable and stable wireless communication performance.

\begin{figure}
\centering
\includegraphics[scale=0.31]{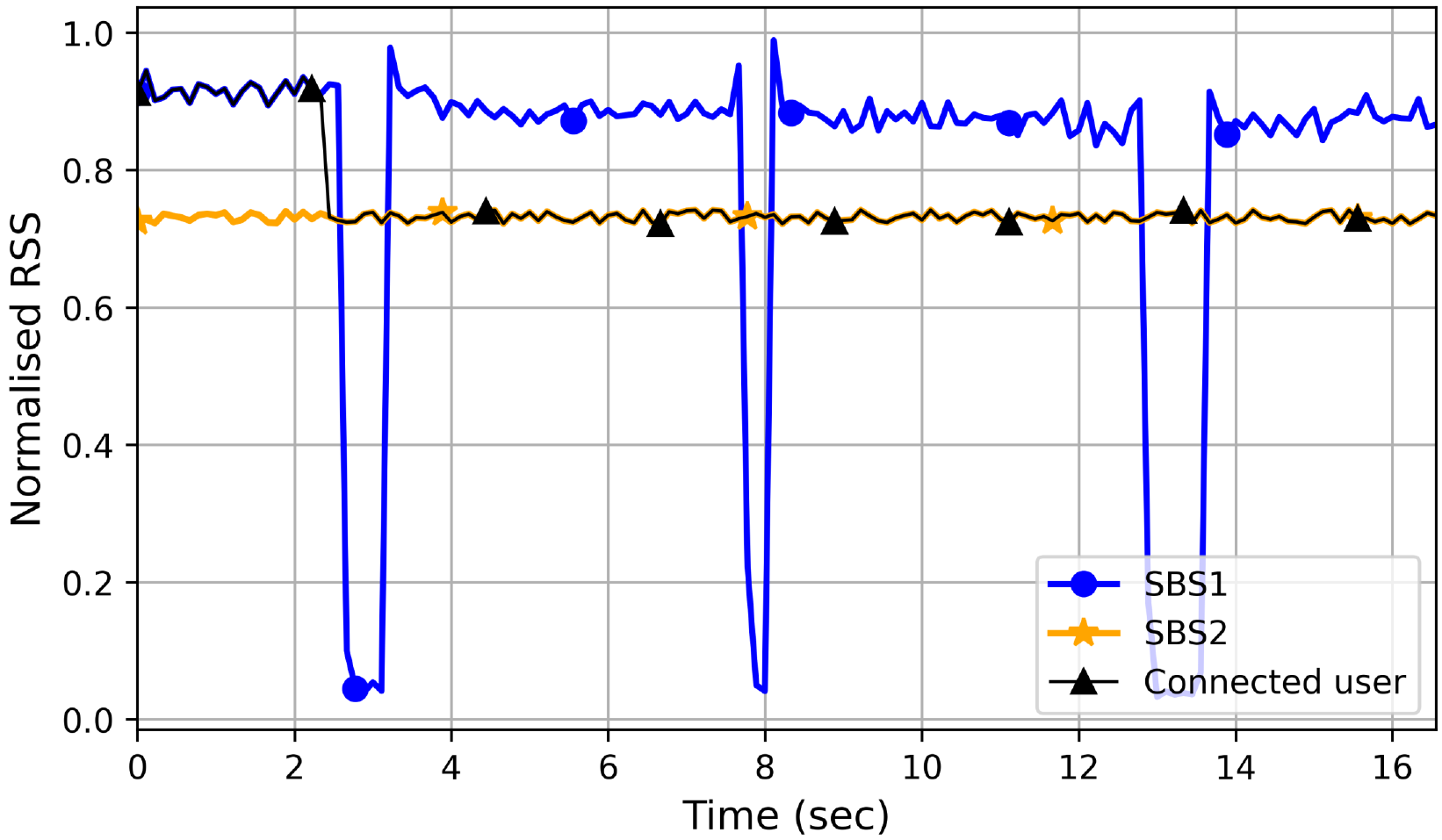}
\caption{The impact of blockages on the user's RSS and how RaDaR is able to detect blockages and ensure seamless connectivity.}
\label{TwoSBSs}
\end{figure}

Next, we examine the efficacy of the proposed RaDaR framework by augmenting the scenario 30 testbed using Wireless InSite ray-tracing software. Specifically, we introduced a new SBS (SBS2) with the same specifications as the existing SBS1, placed 80 metres apart on the opposite side of the street. Fig. \ref{TwoSBSs} depicts the normalised RSS from both SBSs and the SBS that serves the wireless user at different time instances. Once the user is connected, the RaDaR framework begins to function by monitoring the surrounding area using radar sensors. Prior to the first blockage event, the user is connected to SBS1 since it is within the SBS’s coverage area and receives a higher RSS from SBS1 than SBS2. When an obstacle enters the FoV of the radar, the framework detects the object, and the communication system becomes aware of this potential blockage. The framework classifies the object as a blockage and predicts the blockage time. It then determines the optimal time to perform PHO. The figure reveals how the RaDaR framework can detect blocking objects and switch the user to SBS2, which  offers a more stable communication channel. It is essential to note that the QoE of the user when served by SBS1 is better than that when connected to SBS2. However, scarifying the perceived QoE slightly is preferable to experiencing disconnection and engaging in undesired network operation to resume the connection, which impacts network latency and affects its reliability.

\begin{figure}
\centering
\includegraphics[scale=0.33]{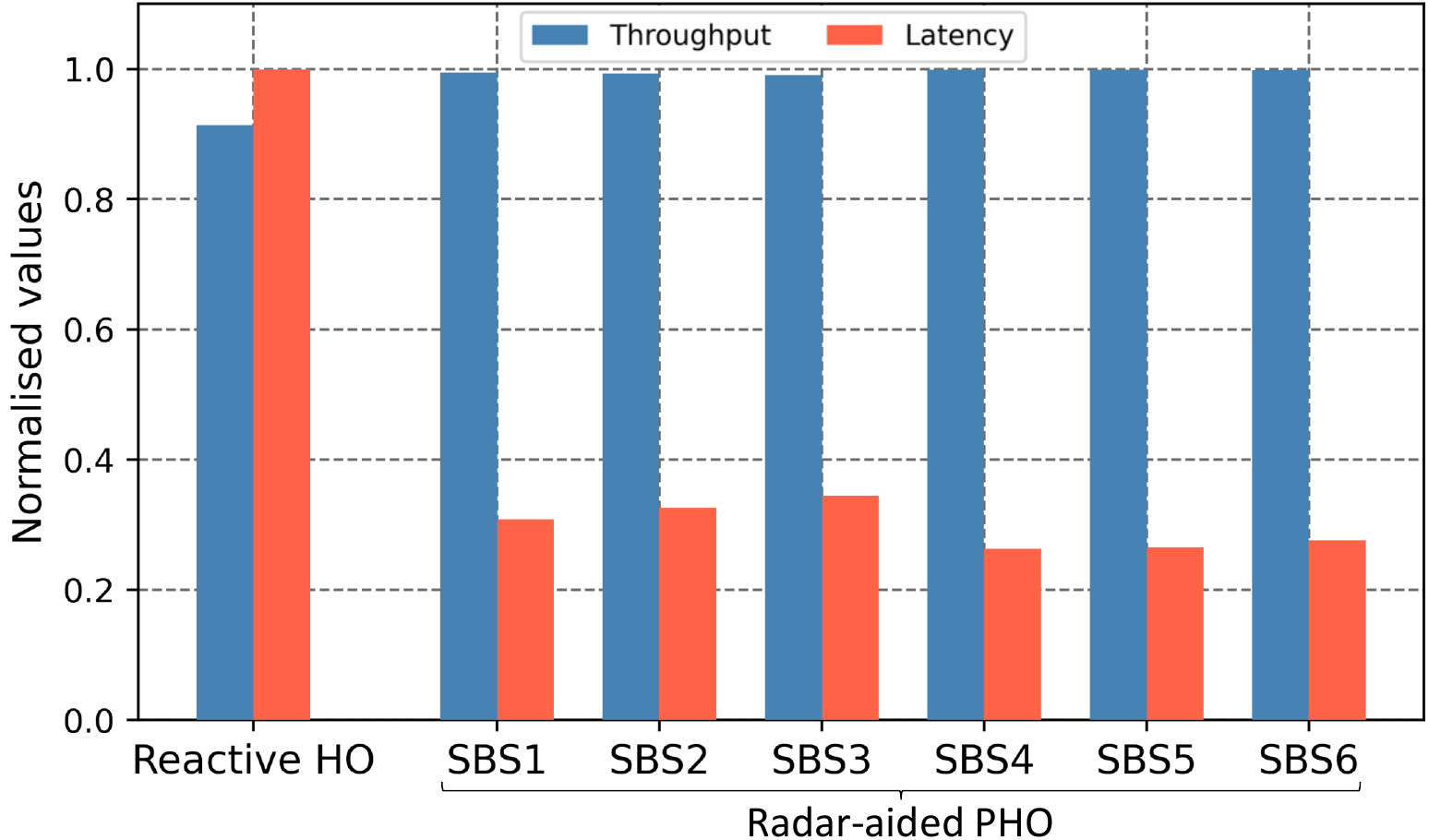}
\caption{Latency and throughput study.}
\label{LT}
\end{figure}

Finally, we examine two key performance metrics for wireless networks, namely the network latency associated with performing handovers and average user throughput for Reactive-HO and radar-aided PHO networks. Our study involves monitoring the evaluation environments over a certain period of time and considering multi-users by treating the stationary user as a new user whenever a new obstacle is detected. Fig. \ref{LT} shows the normalised results of average latency and throughput for both Reactive-HO and radar-aided PHO networks, represented by SBS1-SBS6. In terms of average latency, as discussed in Section \ref{PHOLatency}, Reactive-HO communication systems must perform four steps that take 312.2 ms each time a link is disconnected and a user needs to be switched to another SBS. In contrast, our framework eliminates the first three steps, reducing PHO latency to 80 ms. Making use of $\mathfrak{S}_{PHO}$, the average PHO latency per user is measured as follows:
\begin{equation}
\small
    \zeta=\frac{ \{\mathfrak{S}_{PHO} \times U\}\times 80 + \{(1-\mathfrak{S}_{PHO}) \times U\}\times 312.2}{U}.
\end{equation}
Overall, SBSs adopting the RaDaR framework outperformed Reactive-HO networks lacking proactiveness in detecting blockages. Moreover, the average latency decreased with an increase in $\mathfrak{S}_{PHO}$ as the probability of detecting blockages and performing successful PHO increased. The variations in the average latency values across SBSs are attributed to the differences in the $\mathfrak{S}_{PHO}$, which heavily depend on the serving environment of the SBS.

Regarding average throughput, this study is performed by monitoring the environments for a specific period of time and recording the user’s throughput at every time instant, irrespective of the presence of obstacles. Reactive-HO networks experienced a significant drop in user throughput due to LoS beam disconnection, which required users to switch to other reflected beams with reduced throughput. However, radar-aided PHO systems have a higher probability of predicting obstacles in advance and switching users to an SBS, offering a stable connection, thereby maintaining high throughput levels. 

\section{Conclusions} \label{Conclusion}
In this paper, we proposed a radar-aided dynamic blockage recognition framework called RaDaR. The main objective is to increase high-frequency networks' awareness of their surrounding environment and improve network reliability. We utilised radar measurements for training a dual-output NN model using FL to predict forthcoming link blockages and determine the optimal time to perform PHO, thereby avoiding link disruption. To compare the effectiveness of RaDaR, we considered a conventional wireless network lacking proactive blockage prediction mechanisms, named Reactive-HO. We proceeded to assess the performance of the suggested framework by utilizing co-existing modalities derived from both real-world and simulated environments. The experimental and simulation results confirmed that RaDaR, with its blockage-aware approach, enhances the QoE for users who require low latency and operate in extremely dynamic environments. Compared to Reactive-HO networks, evaluation results indicated that our framework significantly improves the operation of next-generation wireless networks by offering high RSS, maintaining high throughput levels, and reducing network latency, enabling future latency-sensitive applications.

\balance
\bibliographystyle{IEEEtran}
\bibliography{PHO}

\end{document}